\documentclass[prx,twocolumn,showpacs,amsmath,amssymb,superscriptaddress]{revtex4-2}
\usepackage{graphicx}
\usepackage{dcolumn}
\usepackage{bm}
\usepackage{hyperref}
\hypersetup{colorlinks,citecolor=blue, filecolor=blue, linkcolor=blue , urlcolor=blue}

\begin{document}

\title{Magnon gap excitations and spin-entangled optical transition \\ in van der Waals antiferromagnet NiPS$_3$}

\author{Dipankar Jana}
    \email{dipankar.jana@lncmi.cnrs.fr}
    \affiliation{Laboratiore National des Champs Magn\'etiques Intenses, LNCMI-EMFL, CNRS UPR3228,Univ. Grenoble Alpes, Univ. Toulouse, Univ. Toulouse 3, INSA-T, Grenoble and Toulouse, France}

\author{P.~Kapuscinski}
    \affiliation{Laboratiore National des Champs Magn\'etiques Intenses, LNCMI-EMFL, CNRS UPR3228,Univ. Grenoble Alpes, Univ. Toulouse, Univ. Toulouse 3, INSA-T, Grenoble and Toulouse, France}

\author{I.~Mohelsky}
    \affiliation{Laboratiore National des Champs Magn\'etiques Intenses, LNCMI-EMFL, CNRS UPR3228,Univ. Grenoble Alpes, Univ. Toulouse, Univ. Toulouse 3, INSA-T, Grenoble and Toulouse, France}

\author{D.~Vaclavkova}
    \affiliation{Laboratiore National des Champs Magn\'etiques Intenses, LNCMI-EMFL, CNRS UPR3228,Univ. Grenoble Alpes, Univ. Toulouse, Univ. Toulouse 3, INSA-T, Grenoble and Toulouse, France}

\author{I.~Breslavetz}
    \affiliation{Laboratiore National des Champs Magn\'etiques Intenses, LNCMI-EMFL, CNRS UPR3228,Univ. Grenoble Alpes, Univ. Toulouse, Univ. Toulouse 3, INSA-T, Grenoble and Toulouse, France}

\author{M.~Orlita}
    \affiliation{Laboratiore National des Champs Magn\'etiques Intenses, LNCMI-EMFL, CNRS UPR3228,Univ. Grenoble Alpes, Univ. Toulouse, Univ. Toulouse 3, INSA-T, Grenoble and Toulouse, France}
    \affiliation{Institute of Physics, Charles University, Ke Karlovu 5, Prague, 121 16, Czech Republic}

\author{C.~Faugeras}
    \affiliation{Laboratiore National des Champs Magn\'etiques Intenses, LNCMI-EMFL, CNRS UPR3228,Univ. Grenoble Alpes, Univ. Toulouse, Univ. Toulouse 3, INSA-T, Grenoble and Toulouse, France}

\author{M.~Potemski}
    \email{marek.potemski@lncmi.cnrs.fr}
    \affiliation{Laboratiore National des Champs Magn\'etiques Intenses, LNCMI-EMFL, CNRS UPR3228,Univ. Grenoble Alpes, Univ. Toulouse, Univ. Toulouse 3, INSA-T, Grenoble and Toulouse, France}
     \affiliation{CENTERA Labs, Institute of High Pressure Physics, PAS, 01 - 142 Warsaw, Poland}

\begin{abstract}
Optical magneto-spectroscopy methods (Raman scattering, far-infrared transmission, and photoluminescence) have been applied to investigate the properties of the NiPS$_3$ semiconducting antiferromagnet. The fundamental magnon gap excitation in this van der Waals material has been found to be split into two components, in support of the biaxial character of the NiPS$_3$ antiferromagnet.
Photoluminescence measurements in the near-infrared spectral range show that the intriguing $1.475$~eV-excitation unique to the NiPS$_3$ antiferromagnetic phase splits upon the application of the in-plane magnetic field. The observed splitting patterns are correlated with properties of magnon excitations and reproduced with the simple model proposed. Possible routes towards a firm identification of the spin-entangled $1.475$~eV-optical excitation in NiPS$_3$, which can hardly be recognized as a coherent Zhang-Rice exciton, are discussed.

\end{abstract}

\maketitle

\section{Introduction}

Scientific curiosity and the possible design of novel devices continue to drive pertinent research efforts focused on two-dimensional materials \cite{Geim2013, Bhimanapati2015}. Among systems of intense current interest are layered magnets \cite{wang2022, jiang2021} and, in particular, the antiferromagnets from a large family of transition metal phosphorus trichalcogenides (TMPTC),  such as MnPS$_3$, FePS$_3$, NiPS$_3$, CoPS$_3$, MnPSe$_3$, and many others \cite{chaudhuri2022, lanccon2016, liu2021, Vaclavkova2021, pawbake2022, dirnberger2022, Kang2020, Wildes2022, kim2020, Wildes2023}. The TMPTC layers are weakly coupled by van der Waals forces. The magnetic ordering in these antiferromagnets is largely governed by the spin-spin exchange interactions within the layers whereas the interlayer exchange integrals are rather small \cite{Wildes2015, Wildes2022}. This justifies the two-dimensional character of such magnetic systems even in their bulk form. The magnetic anisotropies (of single ions or due to dipolar spin interaction) is another key element that affects the magnetic properties of TMPTC antiferromagnets, establishing the direction of spin ordering and character of magnon excitations (magnon gaps in particular) \cite{Rezende2019, Kim2021}. Two groups of TMPTC antiferromagnets can be distinguished: the group of nearly uniaxial antiferromagnets such as, for example, MnPS$_3$ and FePS$_3$ with spins oriented mostly perpendicular to the layer plane \cite{kobets2009, lanccon2016, liu2021, Vaclavkova2021, pawbake2022} and another group of TMPTC antiferromagnets with spins oriented in the layer planes, which generally represent the biaxial systems (with non-negligible magnetic anisotropy fields along two different axes) \cite{Lancon2018, Wildes2022, Wildes2023, mai2021}. Whereas numerous works have been devoted to reveal the properties of the former-group antiferromagnets, the in-plane TMPTC antiferromagnets are less understood.

Particularly controversial are the properties of NiPS$_3$, both with respect to its characteristic magnon excitations \cite{Lancon2018, Afanasiev2021, Belvin2021, Mehlawat2022, Wildes2022} as well as in regards to the intriguing optical transition in the near-infrared spectral range, which is observed in this material only below N\'{e}el temperature, T$_N$ $\approx 155$~K \cite{ Kim2018, Kang2020, Wang2021}. Bulk NiPS$_3$ in its antiferromagnetic state, in which spins are aligned in the layers’ planes, has been initially considered to be uniaxial, with a characteristic double degenerated fundamental magnon gap \cite{Lancon2018}.
On the other hand, lifting of the magnon gap degeneracy into two components has been evoked in more recent reports,  thus pointing out towards the biaxial character of the NiPS$_3$ antiferromagnet \cite{Afanasiev2021, Wildes2022}. Nevertheless, the identification of these two magnon-gap components is not fully transparent, including a noticeable spread of the reported energy values \cite{Lancon2018, Afanasiev2021, Belvin2021, Mehlawat2022, Wildes2022} and of the associated amplitude of the spin-flop field \cite{Wang2021, Basnet2021, Basnet2022, Bougamha2022}. Perhaps even more confusing is the understanding of a sharp and strongly linearly polarized optical transition, which appears in NiPS$_3$ at $\approx 1.475$~eV only in the antiferromagnetic phase \cite{ Kim2018, Kang2020, Wang2021}. Appealingly enough, this transition has been identified as a collective excitation, the spin-entangled excitonic transition between Zhang-Rice triplet and singlet states \cite{ Kang2020}. In support of such an assignment, this exciton has been claimed to be robust to the application of a magnetic field \cite{ Kang2020}. On the other hand, the axis of its linear polarization has been more recently shown to follow the direction of the Ni$^{2+}$ spins' alignment, which, notably can rotate with the applied in-plane magnetic field \cite{ Wang2021}.  

In this paper, the biaxial character of the NiPS$_3$ antiferromagnet is confirmed via the  observation of two low-energy excitations, at 
$1.25 \pm0.13$~meV ($10 \pm1$~cm$^{-1}$) and $5.33 \pm0.06$~meV ($42.5 \pm0.5$~cm$^{-1}$) as measured at low temperature,
which we identify as two components of the fundamental magnon gap in this antiferromagnet. These excitations soften with temperature and show the evolution with the applied magnetic field as expected for a biaxial antiferromagnet. The analysis of the data provides a set of parameters (such as the ratio of the anisotropy fields, the spin-flop field, and the effective $g$-factor) that characterize the antiferromagnetic phase of NiPS$_3$. 

On the other hand, our examination of the optical response of NiPS$_3$ in the near-infrared spectral range sheds new light on the properties of the attention-catching optical excitation (at $\approx 1.475$~eV) in this antiferromagnet. In striking contrast to recent claims \cite{ Kim2018}, we observe that this transition shows a prominent splitting when the field is applied along the plane directions. The amplitude of this splitting depends on the direction of the applied magnetic field with respect to the spin orientation and it substantially collapses above the spin-flop field. The properties of this $1.475$~eV-excitation are correlated with those of magnon-gap excitations and we propose a simple formalism to ascribe the observed spitting pattern.

\section{Experimental details}
The applied experimental techniques include a variety of optical spectroscopy methods: micro Raman scattering, far-infrared absorption, as well as micro-photoluminescence spectroscopy in the near-infrared spectral range. When needed and feasible, the optical response has been measured as a function of temperature, as a function of the external magnetic field, in different configurations of the direction of the magnetic field with respect to the spin alignment, and also employing polarization resolved techniques. 

The samples used in our experiments were extracted from commercially available NiPS$_3$ crystals. Large area ($\approx 1$~cm$^2$) and rather thick ($\approx 1$~mm) species were used for far-infrared magneto-absorption measurements. Instead, smaller size specimens ($\approx1000$~$\mu$m$^2$ $\times$ 5~$\mu$m) were prepared for Raman scattering and optical measurements in the near-infrared range. Those samples consisted of  rather homogeneous flakes which were mechanically exfoliated from bulk crystals and deposited on a silicon substrate with the help of the dry-transfer technique. More experimental details can be found in the Supplemental Material (SM) \cite{SuppInfo}.

\section{Experimental results and discussion}

\begin{figure}[bt]
	\includegraphics[width=8.4cm]{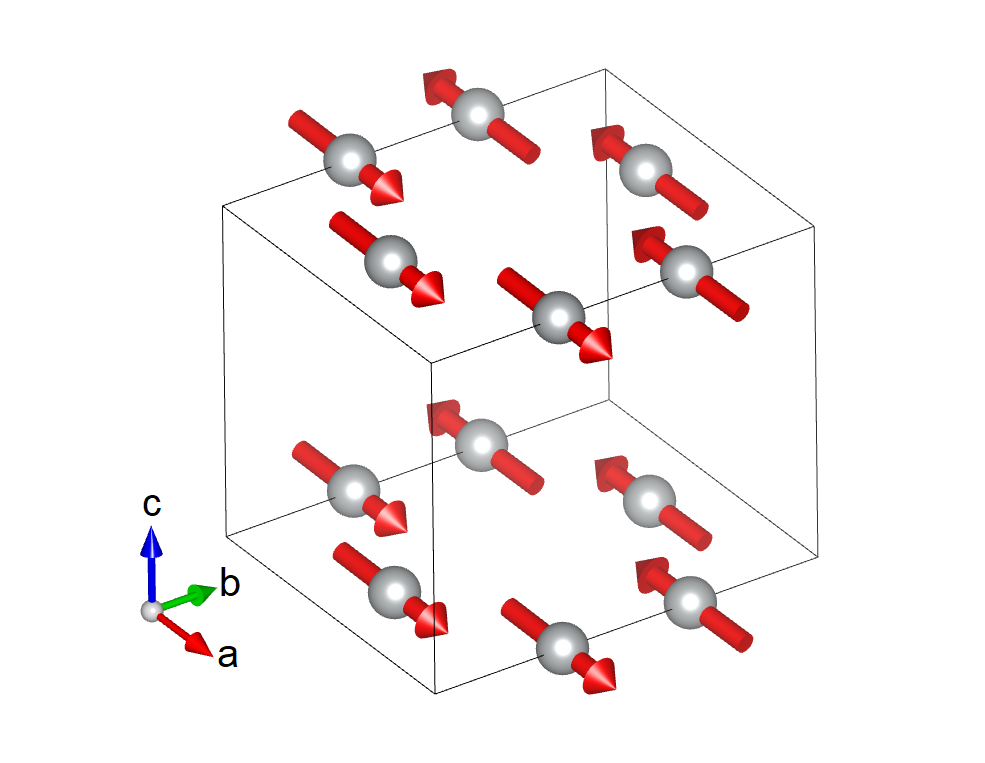}
	\caption{The magnetic structure of NiPS$_3$ in its antiferromagnetic phase. Grey spheres and red arrows represent Ni$^{2+}$ ions and spin direction, respectively. The figure is created using the VESTA software package \cite{ Momma2011}.}
	\label{fig:fig1}
\end{figure}

The magnetic structure of the NiPS$_3$ antiferromagnet, as it emerges from recently reported studies, \cite{Wildes2022} is shown in Fig.~\ref{fig:fig1}. The collinear nickel magnetic moments are aligned in the layers'  \textit{a-b} planes, along the \textit{a}-axis. Such arrangement of Ni$^{2+}$ spins is assumed to be imposed by the anisotropy field along the \textit{c}-axis (which forces the spin alignment in the layer plane) and another, likely weaker, anisotropy field along \textit{b}-axis which aligns spins along the \textit{a}-axis. The net antiferromagnetic spin order is believed to be due to the strong, third-neighbor antiferromagnetic-type exchange coupling between Ni$^{2+}$ spins in the plane. The weaker, nearest-neighbor in-plane and between-the-planes exchange interactions are ferromagnetic. The second neighbor exchange interaction seems to be even weaker.

\subsection{Low-energy spin wave excitations}

\begin{figure}[bt]
	\includegraphics{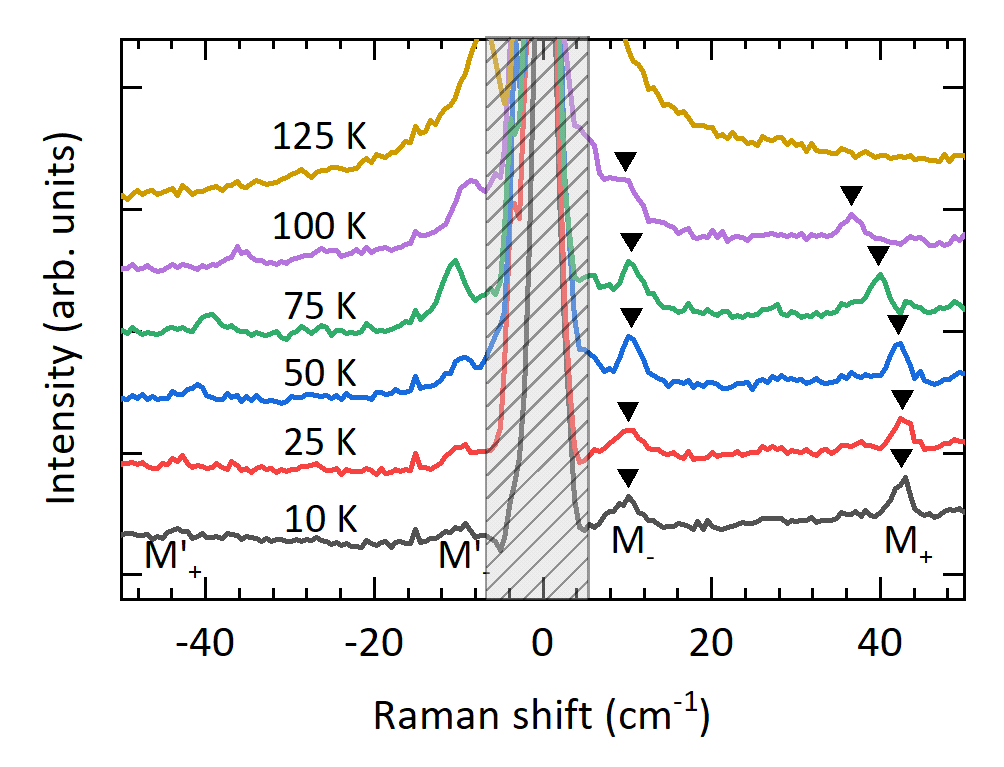}
	\caption{Raman scattering spectra of  NiPS$_3$ antiferromagnet measured at selected temperatures. M$_+$ (M$_+^{'}$) and M$_-$ (M$_-^{'}$) resonances correspond to Stokes (anti-Stokes) modes of the split components of magnon gap excitation. Spectra are shifted vertically for clarity. Peak energies of Stokes modes are marked by black triangles. The shaded region marks the spectral range blocked by the Bragg filters.}
	\label{fig:fig2}
\end{figure}

The presence of two anisotropy fields, i.e., the biaxial character of the NiPS$_3$ antiferromagnet, imposes the splitting of its fundamental magnon gap into two components. As shown below, these two low-energy spin-wave excitations (magnons) at the $k=0$ point of the Brillouin zone, can be traced with Raman scattering experiments, and the upper energy mode is also visualized with far-infrared transmission measurements. While the Raman scattering technique has been already applied to study NiPS$_3$ specimens, only the strong response due to the characteristic phonon modes that appear at relatively high energies have been investigated so far \cite{Wang2021, kim2020, Kim2018}. Instead, in our experiments we concentrate on the Raman scattering response in the close vicinity, $\pm~50$ cm$^{-1}$, of the laser line. As shown in Fig.~\ref{fig:fig2}, two distinct excitations are observed in this spectral range. They are labeled as M$_+$ and M$_-$ in the Stokes Raman scattering signal and have their M$_+^{'}$ and M$_-^{'}$ counterparts in the anti-Stokes spectra. It is worth noticing that Raman scattering features related to our low-energy modes are significantly less intense than those due to phonons observed at higher energies (see Fig. S1 of SM) \cite{SuppInfo}. 
From the data illustrated in Fig.~\ref{fig:fig2} we read that in the limit of low temperatures the characteristic energies of our M$_{+/-}$ excitations are, respectively, $w_{M_+} \cong 43$~cm$^{-1}$ and $w_{M_-} \cong 10$~cm$^{-1}$. These energies decrease when the temperature is raised; the effect being particularly well visible for the case of the upper energy M$_+$ mode. When approaching the Ne\'{e}l temperature the M$_+$ significantly broadens and the M$_-$ mode merges into the enlarged laser tail. The observed, temperature-activated softening of M$_{+/-}$ modes stands for the primary indication that they are associated with the magnon gap excitations.

\begin{figure*}[htp]
\centering
\includegraphics[width=5.9cm]{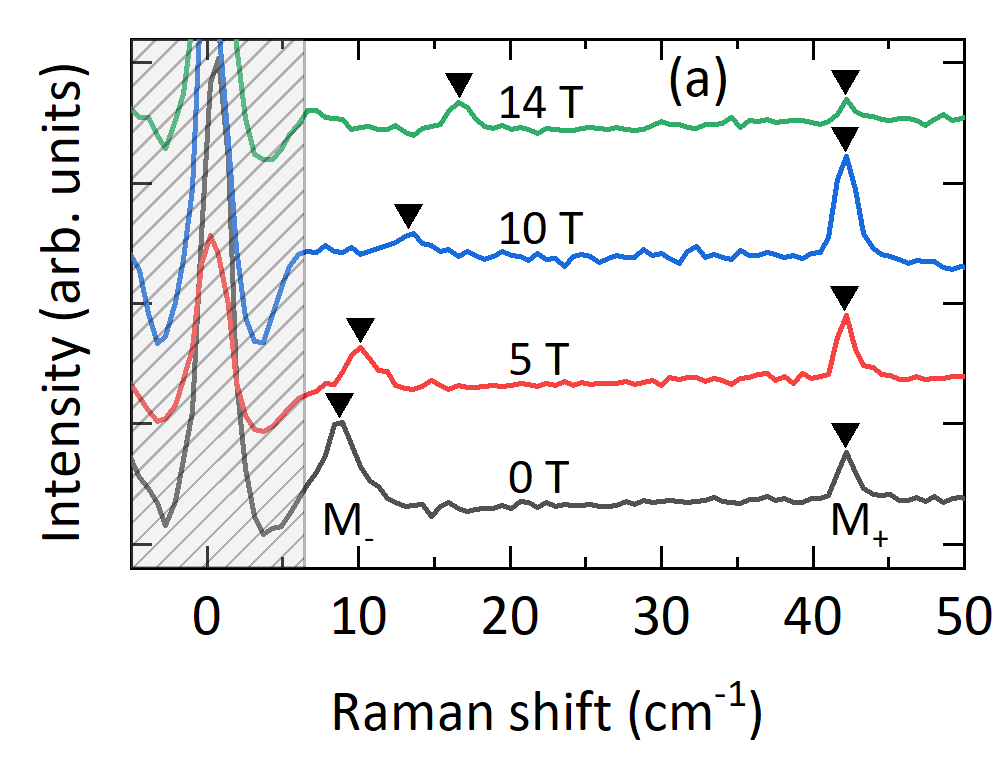}
\includegraphics[width=5.9cm]{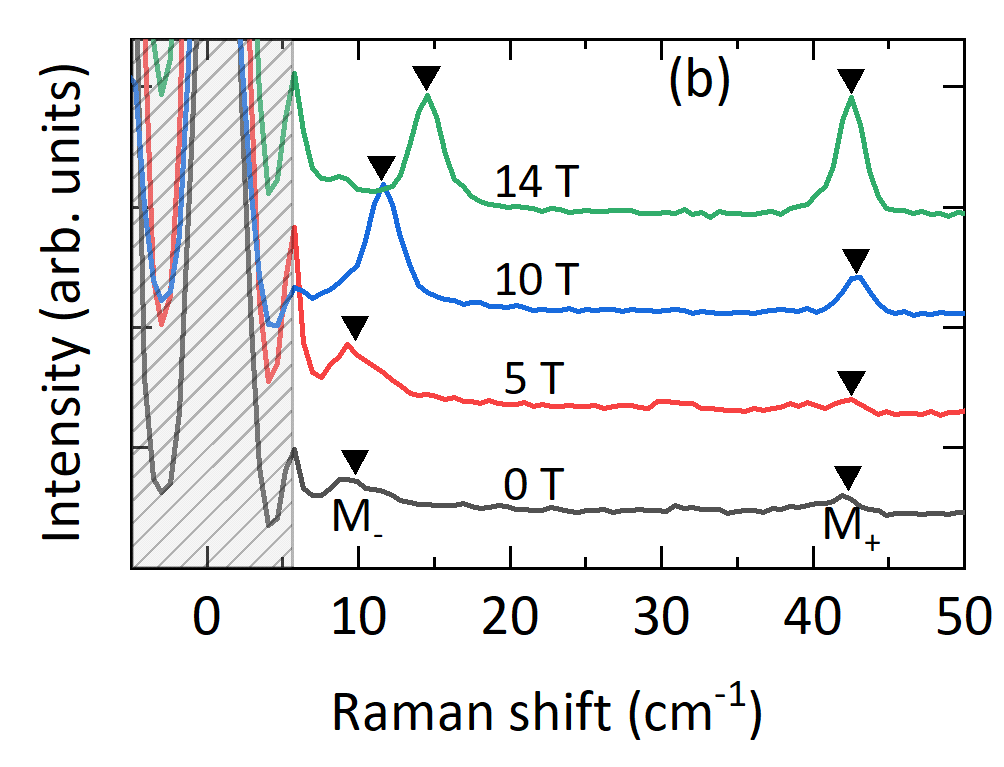}
\includegraphics[width=5.9cm]{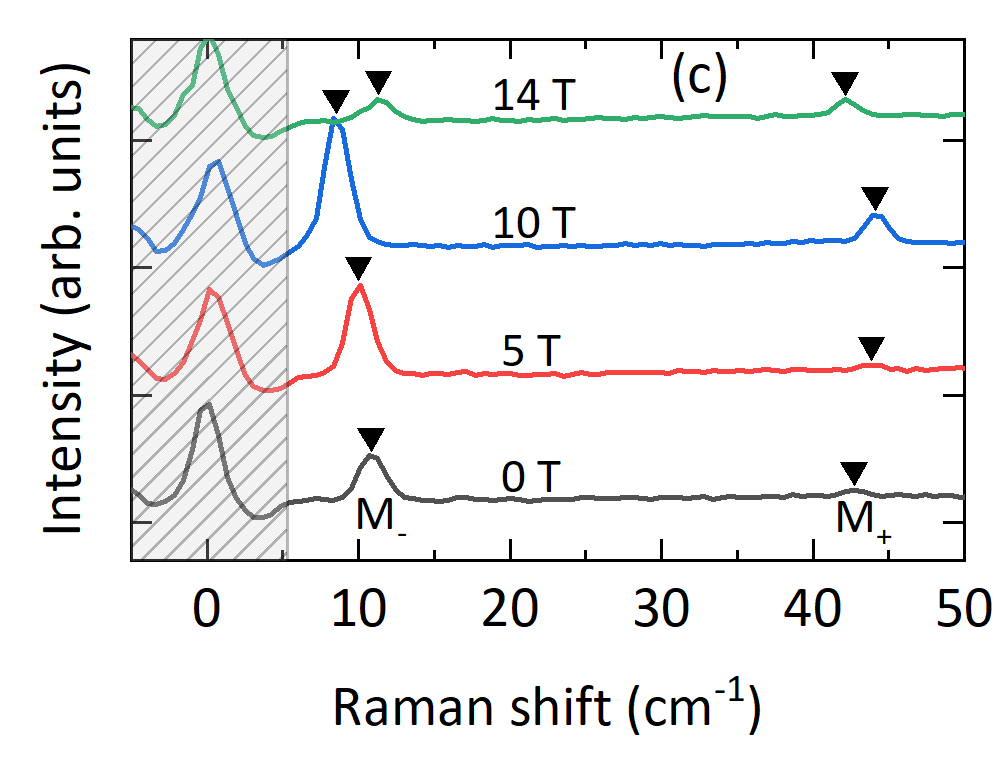}
\caption{(a), (b) and  (c) Low temperature(4.2~K) Raman scattering spectra of exfoliated NiPS$_3$ flakes in three different orientations of the crystal a-axis (i.e. with respect to spin alignment at $B=0$) with respect to the in-plane magnetic field. Magnon gap excitations (M$_+$ and M$_-$)  are marked by a black triangle. Spectra are shifted vertically for clarity. The shaded region marks the spectral range blocked by the Bragg filters.}
\label{fig:Fig3}
\end{figure*}

To further elucidate the origin of our M$_{+/-}$ modes as two low-energy magnon gap excitations, we examine  their evolution upon the applied magnetic field, as investigated at low, $T=4.2$~K, temperature. First, we consider the configuration of the magnetic field applied in the direction along the layer plane. The data obtained for three different NiPS$_3$ flakes are presented in Fig.~\ref{fig:Fig3}. As can be seen in this figure, there is a perceivable evolution of the M$_{+/-}$ modes with the applied magnetic field, albeit it is clearly different for each set of the data. It is logical to expect that the magnetic field evolution of the M$_{+/-}$ modes might be critically altered by the actual experimental geometry, mainly the magnetic field direction with respect to the direction of the spin alignment (at $B=0$~T), i.e., with respect to the \textit{a}-axis of the NiPS$_3$ crystal. Although the orientation of the \textit{a,b} crystal axes with respect to the direction of the applied magnetic field has not been independently determined in our work, we found it to be different in the three flakes investigated.

To interpret the observed dependencies of M$_{+/-}$ energies on the strength of the applied in-plane magnetic field ($B=\mu_0H$) we refer to the mean-field theory of antiferromagnetism \cite{Nagamiya1955}, according to which the $w_{+/-}=w_{+/-}(B)$ dependencies of the magnon gap energies in a simple biaxial antiferromagnet are given by two (positive) solutions of the following, $F(w)=0$, equation: 

\begin{equation} \label{eq:1}
\begin{split}
& F(w)=\left(\frac{w}{g\mu_B}\right)^4-\left(\frac{w}{g\mu_B}\right)^2\left[B^2\left(\cos^2{\Psi+1}\right) +C_2 \right. \\
& \left. +C_1\left\{\cos^2{\left(\Psi-\theta_B\right)}-2\sin^2{\left(\Psi-\theta_H\right)}\right\}\right] +B^4\cos^2{\Psi} \\
&-B^2\left[C_1\left\{\cos^2{\Psi}\cos{2\left(\Psi-\theta_B\right)} \right.\right. \\
& \left.+\cos{\Psi}\sin{\theta_B}\sin{\left(\Psi-\theta_B\right)}+\sin{\Psi}\cos{\theta_B}\sin{\left(\Psi-\theta_B\right)}\right\} \\
& \left. +C_2\left(\cos^2{\Psi}-\sin^2{\Psi}\right)\right] \\
& +C_1\cos{2\left(\Psi-\theta_B\right)}\times\left[C_2-C_1\sin^2{\left(\Psi-\theta_B\right)}\right]  =0
\end{split}
\end{equation}
where $g$ is the effective $g$-factor and $\mu_B$ stands for the Bohr magneton. $\theta_B$ is the angle between the direction of the applied magnetic field and the initial (at $B=0$~T) spin alignment direction (along \textit{a}-axis). The constants $C_1$ and $C_2$ are given by
\begin{equation} \label{eq:2}
C_1=2B_JB_{D-b},\ {\ C}_2=2B_JB_{D-c}\     
\end{equation}
Here, $B_J$, $B_{D-b}$  and $B_{D-c}$ are, respectively, the effective exchange field, and the anisotropy fields along the easy \textit{b}- and hard \textit{c}-axis. As illustrated in Fig. S2 of SM, \cite{SuppInfo}  when $\theta_B \neq 0$, the application of the external magnetic field progressively rotates the magnetic moments towards the direction perpendicular to the field, while the moment directions of two spin sublattices keep being opposite in a wide range of the magnetic field, due to strong exchange coupling. Canting of spins, towards their unavoidable ferromagnetic alignment (along the magnetic field) is likely apparent in the range of very high magnetic field, hardly available in our experiments.

The apparent rotation, $\Psi$, of the spins' alignment with respect to the magnetic field direction is given by the following equation \cite{Nagamiya1955}:

\begin{equation} \label{eq:3}
\tan 2\Psi=\frac{\sin 2\theta_B}{\cos 2\theta_B-B^2⁄B_{sf}^2}
\end{equation}
where, $B_{sf}$ is the spin-flop field. When setting $B=0$ ($\Psi=\theta_B$),  the two roots of Eq.~\ref{eq:1} are:

$w_{M-}(B=0)= g\mu_B\sqrt{C_1}=g\mu_B\sqrt{2B_JB_{D-b}}$

$w_{M+}(B=0)= g\mu_B\sqrt{C_2}=g\mu_B\sqrt{2B_JB_{D-c}}$

which account for the energies of two non-degenerate low-energy magnon modes that appear in a biaxial antiferromagnet in the absence of the magnetic field.

The two, $w_{M+}(B)$ and $w_{M-}(B)$ solutions of Eq.~\ref{eq:1} for several, different values of $\theta_B$ are illustrated in Fig. S3 of SM \cite{SuppInfo}. Notably, if $\theta_B=0$ then in the range of $B<B_{sf}$: 

$w_{M+}= g\mu_B\sqrt{C_2+3B^2}$

$w_{M-}= g\mu_B\sqrt{C_1-B^2}$

and we note that $w_{M-}(B)$ reaches its zero value at the spin-flop field $B_{sf}=\sqrt{C_1} $.
 In this particular case, the arrangement of the Ni$^{2+}$ magnetic moments is such that they remain aligned along the \textit{a}-axis until $B_{sf}$ is reached, at which spins flop abruptly to be aligned in the direction perpendicular to the external field, (i.e. along \textit{b}-axis) though conserving the antiferromagnetic order. 

\begin{figure}[bt]
	\includegraphics{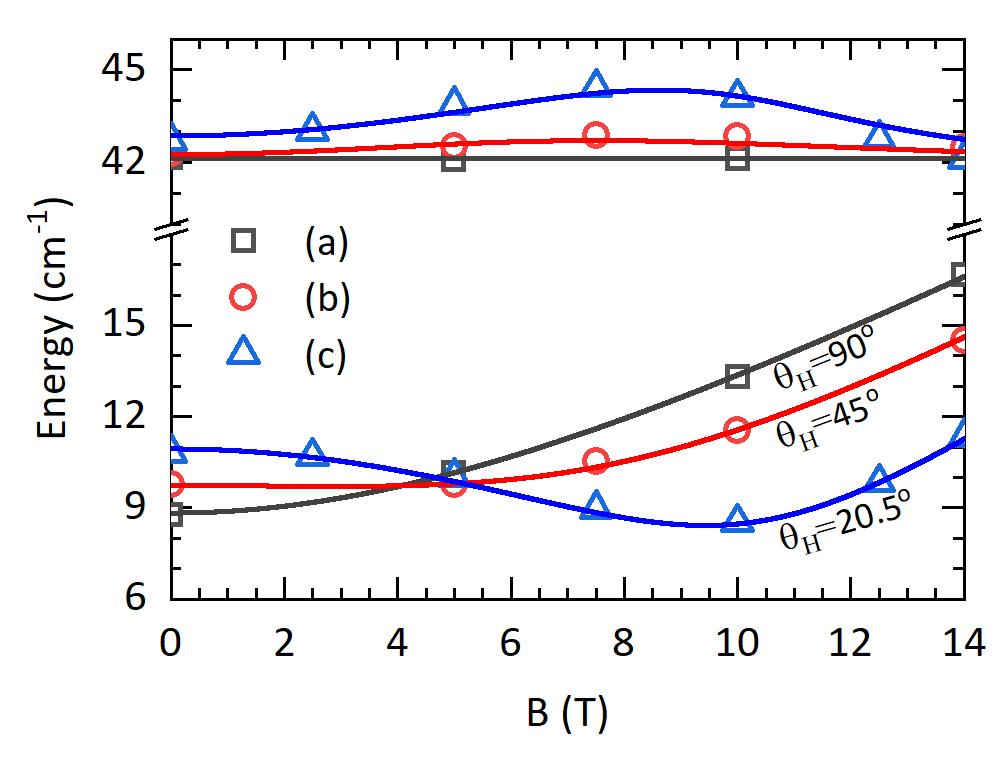}
	\caption{In-plane magnetic field dependence of M$_+$ and M$_-$ magnon excitation energies at 4.5~K as extracted from the results of Raman scattering measurements, illustrated in Fig.~\ref{fig:Fig3}. The simulated field dependence for $\theta_B={90}^{\circ}, {45}^{\circ}$\ and\ ${20.5}^{\circ}$ are also shown by solid lines on the same graph. The same $g$-factor ($g=2.15$) has been considered for three flakes.}
\label{fig:Fig4}
\end{figure}
  
 The energies of $M_+$ and $M_-$ excitations, extracted from the results of magneto-Raman scattering measurements, illustrated in Fig.~\ref{fig:Fig3}, for three different NiPS$_3$ flakes, are shown in Fig.~\ref{fig:Fig4}. The three pairs of $w_{M+}(B)$ and $w_{M-}(B)$ dependencies are clearly different what points out different experimental geometries (different $\theta_B$ angle) for each of three data sets.  Solid lines in Fig.~\ref{fig:Fig4} represent the fitted dependencies according to Eq.~\ref{eq:1}. The simulation of each pair of $w_{M+}(B)$ and $w_{M-}(B)$ dependencies implies the use of four fitting parameters: $g$ (the effective $g$-factor), $C_1$ (or the related energy $w_{M-}=g\mu_B\sqrt{C_1}$), C2 (or $w_{M+}=g\mu_B\sqrt{C_2}$) and the angle $\theta_B$. 
 As expected the three extracted values of 
 $\theta_B={90}^{\circ}, {45}^{\circ}$ and ${20.5}^{\circ}$ are significantly different. The $g$-factor, $g=2.15$ has been found to be common for three flakes. The 
 magnon energies at zero magnetic field are somewhat scattered: $(w_{M-},w_{M+}) = (9~cm^{-1}, 42~cm^{-1}), (10~cm^{-1}, 42~cm^{-1})$ and $(11~cm^{-1}, 42.8~cm^{-1})$, respectively for (a), (b) and (c) data sets (see Fig.~\ref{fig:Fig4}). Slightly different magnon energies for our three flakes may be due to their different thickness and/or the induced strain \cite{Lancon2018}. Setting  $w_{M-}=10$~cm$^{-1}$ and $w_{M+}=42.5$~cm$^{-1}$ we derive $B_{sf}=w_{M-}/g\mu_B=10.4$~T for the spin-flop field and $B_{D-c}/B_{D-b}=(w_{M+}/w_{M-})^2$ = 18 for the ratio of the hard- to easy-axis anisotropy fields.  

 Turning now the attention to the configuration of the magnetic field applied perpendicularly to the layers' planes ($B \parallel c$) we note that in this case all spins, independent of their in-plane orientation, are aligned perpendicularly to the direction of the B-field. When $B \parallel c$, the magnon gap energies versus the magnetic field are given by simple formulas \cite{Nagamiya1955}:
 \begin{equation} \label{eq:4}
w_{M-}= g\mu_B\sqrt{C_1}\qquad w_{M+}= g\mu_B\sqrt{C_2+B^2} 
\end{equation}
The $M_-$ mode is not affected by the magnetic field whereas $w_{M+}$ increases monotonically with $B$.

\begin{figure}[bt]
	\includegraphics{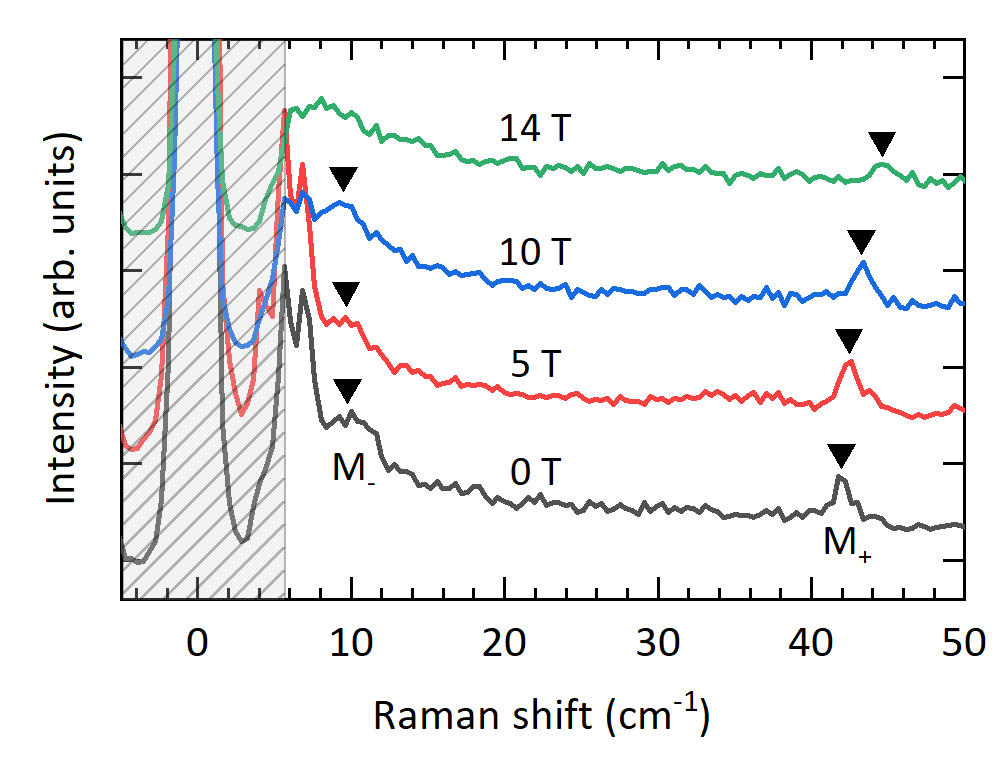}
	\caption{Low temperature (4.2~K) Raman scattering spectra of exfoliated NiPS$_3$ flake at selected strength of the magnetic field applied in the direction tilted by 20$^{\circ}$ with respect to the crystal c-axis. The spectra are shifted vertically for clarity. Peak energies of magnon excitation modes are marked by the black triangle. The shaded region marks the spectral range blocked by the Bragg filters.}
	\label{fig:Fig5}
\end{figure}

\begin{figure}[bt]
	\includegraphics{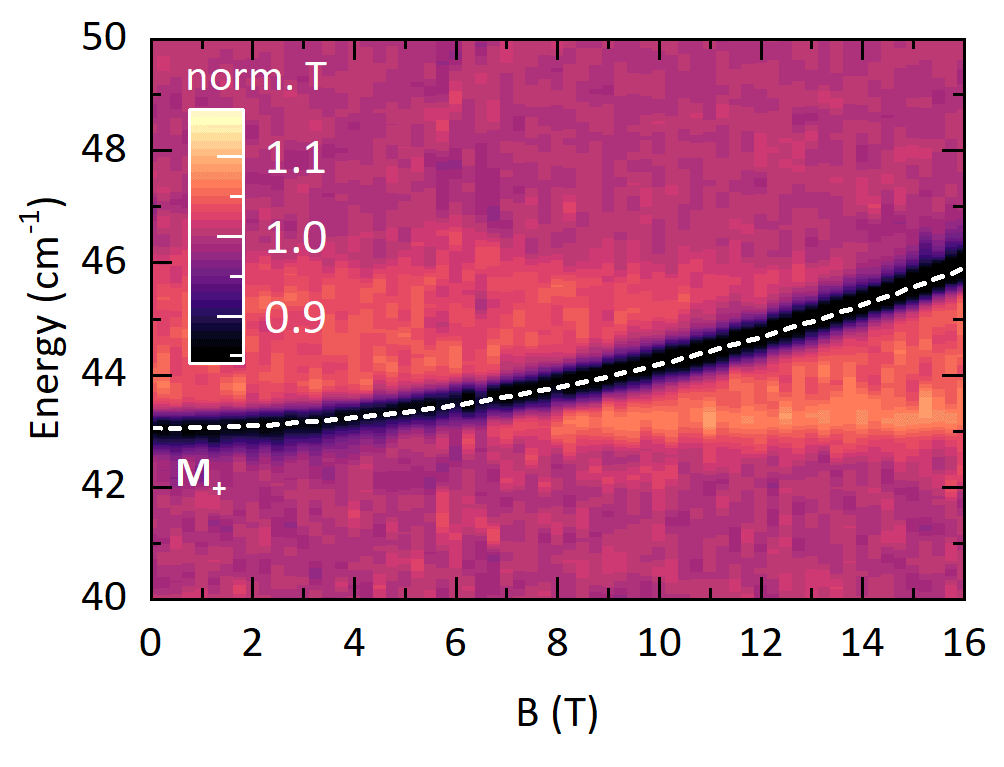}
	\caption{False color map of normalized far-infrared transmission of NiPS$_3$, measured at 4.2~K, as a function of the magnetic field applied along the crystal c-axis. Dashed line corresponds to fit a of the $w_{M+} (B)$ dependence according to Eq. \ref{eq:4}.}
	\label{fig:Fig6}
\end{figure}

When setting the $B \parallel c$ configuration, the performance of our magneto-Raman scattering set-up appeared to be not sufficient enough for measurements of low-energy excitations. The encountered problems to efficiently reduce the stray laser light could be, however, overcome when tilting the sample, i.e., when setting a certain angle $\angle (B,c)$ between the direction of the magnetic field and the crystal c-axis. The magneto-Raman scattering spectra measured in the tilted, $\angle (B,c)= 20^{\circ}$ configuration are illustrated in Fig.~\ref{fig:Fig5}. In such a configuration the dispersion of $M_-$ and $M_+$ modes with the magnetic field cannot be expected to exactly follow the predictions otherwise valid for the ideal case of $B \parallel c$ geometry. Nevertheless, the overall trends are in agreement with Eq.~\ref{eq:4}: $w_{M+}$ increases monotonically with $B$ whereas the $M_-$ mode is unaffected by the magnetic field.

Another possible method to trace the magnon gap excitations in antiferromagnets is a direct measurement of light absorption/transmission associated with these modes in the far infrared (FIR) spectral range \cite{Vaclavkova2021}. The advantage of this technique, in conjunction with the application of the magnetic field, is that it can be easily set in the $B \parallel c$ configuration (Faraday geometry). We must, however, admit that our large NiPS$_3$ specimens, which are imperative for FIR transmission experiments, are composed of differently oriented, in the ($a-b$)-plane, crystal grains. This fact does not affect the analysis of the magneto-spectroscopy data in the $B \parallel c$ geometry (all Ni$^{2+}$ spins are aligned perpendicularly to $B$), but makes FIR magneto-spectroscopy hardly applicable in the configuration of $B$ applied in the ($a-b$)-plane (magnon mode dispersion with $B$ depends critically on actual, $B$ versus $a$-axis orientation). The results of FIR magneto-transmission measurements carried out at low temperature (4.2~K) in the Faraday geometry on bulk NiPS$_3$ samples  are illustrated in Fig.~\ref{fig:Fig6}. A clear absorption feature observed in these data, along with its characteristic evolution with the magnetic field, is identified as due to the $M_+$ excitation. Unfortunately, the detection of the low-energy $M_-$ mode was beyond the limit of our experimental setup. Using the adequate, for $B \parallel c$ geometry, formula (see Eq.~\ref{eq:4}), the $w_{M+}(B)$ dependence is simulated (dashed line in Fig.~\ref{fig:Fig6}) when setting $w_{M+}=43$~cm$^{-1}$ and $g=2.15$, the latter value is in perfect agreement with the estimation of this parameter from magneto-Raman scattering experiments.

\subsection{Spin entangled optical excitation}
\begin{figure}[bt]
	\includegraphics{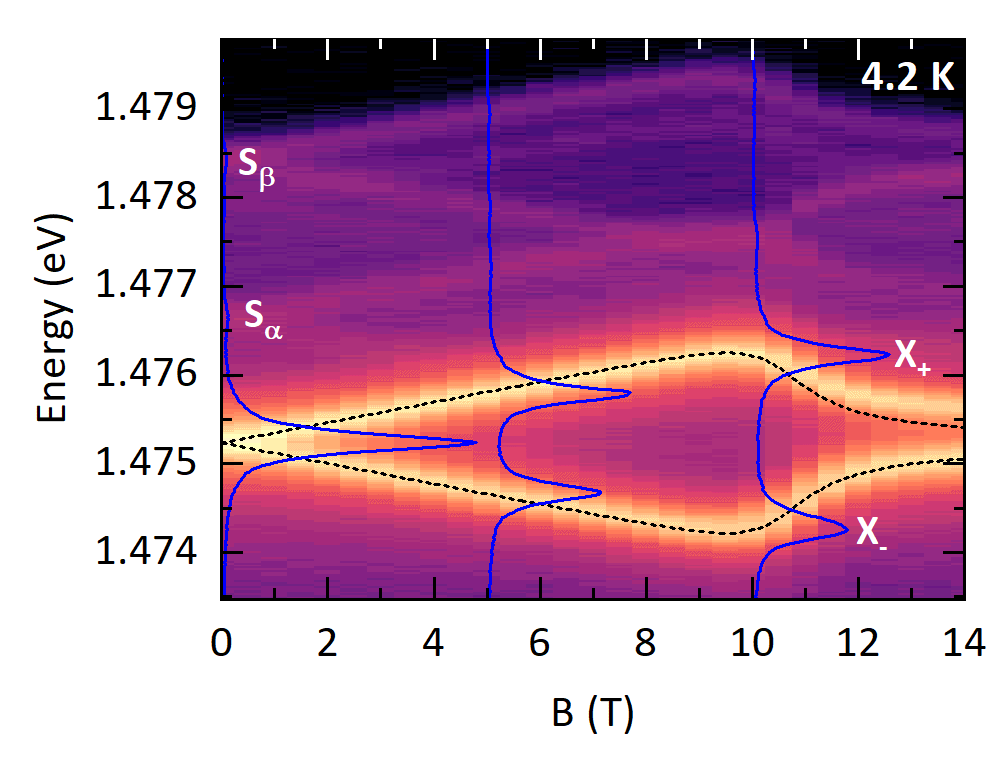}
	\caption{False color map of low temperature (4.2~K) PL of NiPS$_3$ exfoliated flake as a function of the in-plane magnetic field. Few representative PL spectra, measured at a magnetic field strength of 0~T, 5~T, and 10~T are also plotted. Dashed lines correspond to a fit of the $E_{X_{+/-}}$ dependence according to Eq. \ref{eq:5} for $g=2.0$, B$_{sf}$=10.55~T and $\theta$$_B$ = 5$^{\circ}$. High energy satellite peaks are labeled as S$_\alpha$ and S$_\beta$.}
	\label{fig:Fig7}
\end{figure}

\begin{figure*}[htp]
\centering
\includegraphics[width=5.9cm]{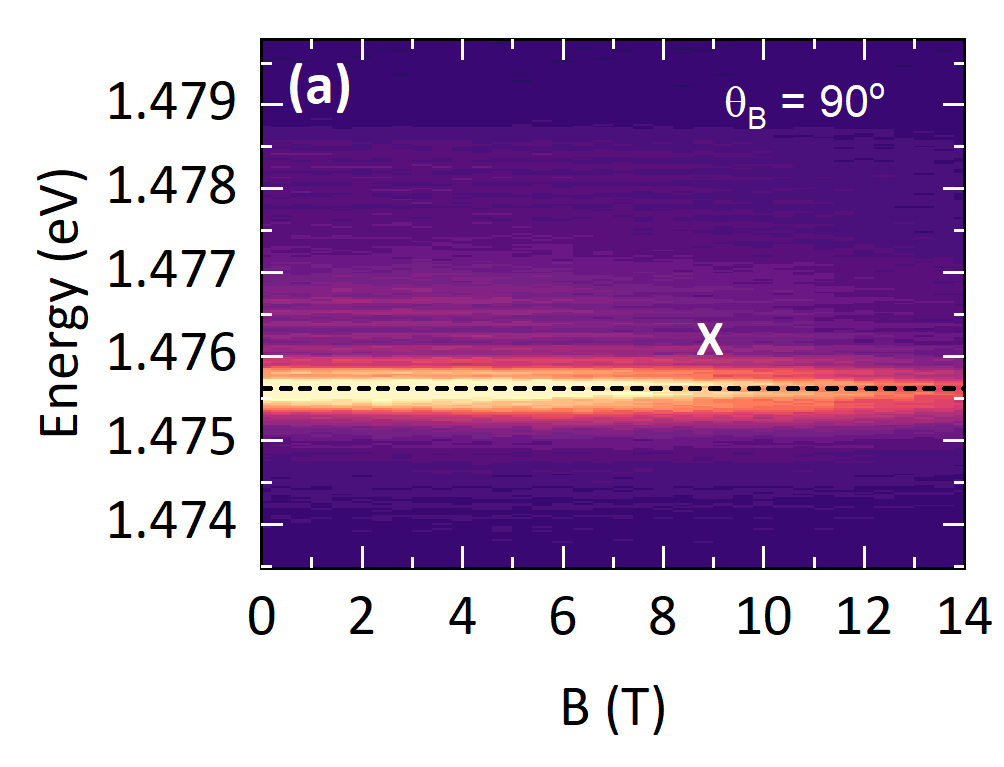}
\includegraphics[width=5.9cm]{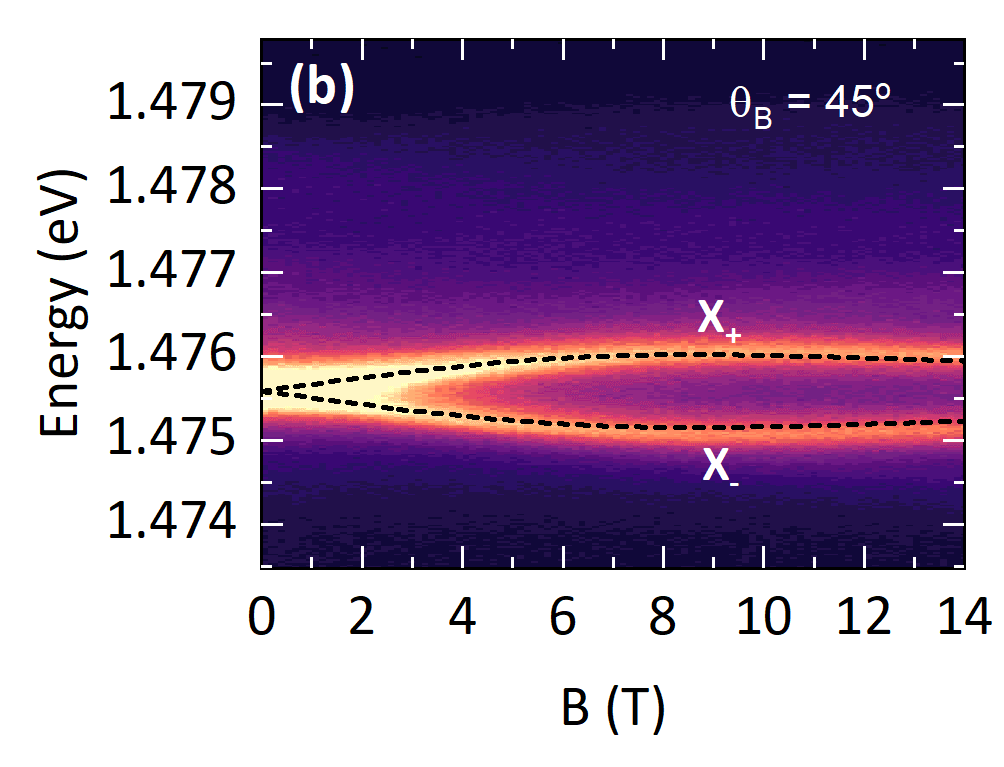}
\includegraphics[width=5.9cm]{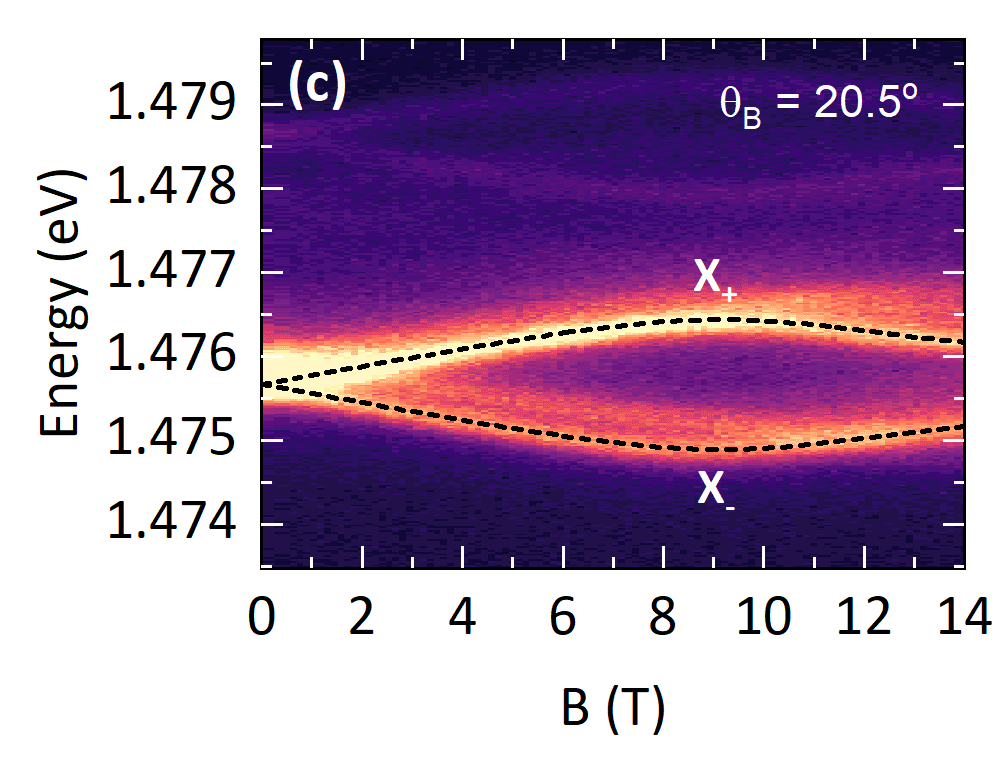}
\caption{(a), (b) and  (c) False color map of low temperature (4.2~K) PL of NiPS$_3$ exfoliated flakes as a function of the in-plane magnetic field with identical spin orientations as in  Fig.~\ref{fig:Fig3}. Dashed lines correspond to a fit of the $E_{X_{+/-}}$ dependence according to Eq. \ref{eq:5} for $g=2.0$.}
\label{fig:Fig8}
\end{figure*}

The optical absorption edge in NiPS$_3$ is located at $\sim 1.8$ eV but this crystal is also known to exhibit several optical transitions below its optical bandgap \cite{Kim2018}. Particularly intriguing are the properties of an exceptionally sharp transition which gives rise to both the photoluminescence (PL) and absorption-type signal at an energy of $\approx 1.475$ eV. As already reported \cite{Wang2021} and confirmed in our study (see Fig. S4 of SM) \cite{SuppInfo}, this transition, referred to here as X-transition, is sensitive to spin ordering: it appears only in the antiferromagnetic phase and displays a high degree of linear polarization, the axis of which follows the direction of spin alignment. The X-transition has been previously identified as a many-body-in-nature bosonic excitation/exciton involving the Zhang-Rice triplet and singlet states \cite{Kang2020}. In support of this concept, the X-transition has been reported to be robust with respect to the applied magnetic field. That is, however, in contradiction with the results of the present experiments.

A remarkable evolution of the X-transition under the applied field is illustrated in Fig.~\ref{fig:Fig7} with a representative collection of the PL spectra measured, at low temperature (4.2~K), as a function of the in-plane magnetic field. In the absence of the magnetic field our spectra perfectly match those reported previously: a sharp ($\leq 0.2$~meV halfwidth) PL peak due to X-transition is observed slightly above $1.475$ eV and it is followed by two/three much weaker satellites on its high energy side. Yet, the application of the in-plane magnetic field induces a splitting of the X-transition (as well as of its high energy satellites) into two, X$_+$ and X$_-$, components (see Fig.~\ref{fig:Fig7}). The separation $\Delta E$ between the X$_+$ and X$_-$, components first increases linearly with the magnetic field but abruptly decreases when $B\ge 10$ T, i.e., above the spin-flop field. The splitting pattern of the X-transition such as that shown in Fig.~\ref{fig:Fig7}  is, however, not the same for all investigated NiPS$_3$ flakes. We speculate that the actual orientation of the magnetic field with respect to the direction of the spin alignment decides about the character of the X-transition splitting. To check this hypothesis, the reported above magneto-Raman scattering measurements on three different flakes (with different $B$ versus \textit{a}-axis orientations) have been in parallel completed by magneto-PL measurements of the X-transition. The data presented in Fig.~\ref{fig:Fig3} (a), (b), (c) and the results presented in Fig.~\ref{fig:Fig8} (a), (b), (c) are, respectively, obtained for the same NiPS$_3$ flakes. As shown in Fig.~\ref{fig:Fig8} (a), the splitting of the X-transition is practically absent in the configuration when the magnetic field is applied perpendicularly to the direction of the Ni$^{2+}$ spins' alignment ($\theta_B = 90^\circ$ and thus $\Psi(B)= 90^\circ$, see Fig. S3a of SM \cite{SuppInfo}) but progressively appears when $\theta_B$ becomes smaller (see Fig.~\ref{fig:Fig8} (b), and (c)). The observed nonlinearities in the X-transition splitting must be due to the field-induced rotation of the axis of Ni$^{2+}$ spin alignment (see Fig. S3 (a) of SM) \cite{SuppInfo}. The splitting of the X-transition is therefore expected to reflect the field-induced disequilibrium of magnetic moments of Ni$^{2+}$ spin sublattices, one of the moments being enhanced and another suppressed along the direction of the applied magnetic field.
In other words, we propose that the observed splitting of the X-transition can be accounted for by the simple following formula:
\begin{equation} \label{eq:5}
E_{X_{+/-}} = E_{X_0} \pm g\mu_B B \cos(\Psi)
\end{equation}
where $\Psi=\Psi(B)$ is the previously defined angle (Eq.~\ref{eq:3}) between the axis of the Ni$^{2+}$ spins' alignment and $B$-direction; $g$ stands for the effective $g$-factor and $E_{X_0}$ is the energy of the X-transition at zero magnetic field.
As shown with dashed lines, in Fig.~\ref{fig:Fig7} and Fig.~\ref{fig:Fig8} (a), (b) and (c), the above formula reproduces well the observed splitting patterns of the X-transition. The results were simulated assuming $g=2.0$ for all data sets. The $E_{X_0}$ energy has been found to scatter a bit from $1.4752$ to $1.4755$, depending on the flake investigated. When simulating the $E_{X_{+/-}}(B)$ dependencies for the data shown in Fig.~\ref{fig:Fig8} (a), (b) and (c) we have, respectively, set $\theta_B= 90^\circ, 45^\circ$ and $20.5^\circ$, as previously derived from the analysis of magnon gap excitations.
As for results presented in Fig.~\ref{fig:Fig7} (which were not completed by Raman scattering measurements) the dashed lines follow the formula with the best-fit parameters $\theta_B= 5^\circ$
and spin flop field $B_{sf}=10.55$ T. 
It is important to note that our measurements confirm a strong linear polarization of the PL spectra related to X-transition. Moreover, we observe that the polarization properties are the same for both $X_+$ and $X_-$ components and the polarization axis rotates with the applied, in-plane magnetic field. The results of polarization measurements of spectra illustrated in Fig.~\ref{fig:Fig7} are displayed in Fig. S4 of the SI. In this case, the field-induced gradual rotation of the polarization axis is rather weak until reaching the spin-flop field ($B_{sf}\sim 10$ T) where it shows an abrupt change. This behavior is similar to the field dependence of $\Psi$ (see Fig. S2 and S3a of the SM) \cite{SuppInfo}, which reveals a correlation of the polarization axis of the X-transition with the direction of spin alignment.

We do believe that our study reveals the overlooked but relevant properties of the X-transition in NiPS$_3$, which recently attracted considerable attention. The firm identification of this intriguing excitation calls for thorough theoretical analysis and is beyond the scope of our experimental work. We should, however, admit that the reported here sensitivity of the X-transition to the applied magnetic field (splitting of the transition, in particular) is in clear disagreement with the statement on the robustness of this transition to the magnetic field; the statement being an argument to assign the X-transition as due to a coherent exciton involving the Zhang-Rice triplet and singlet states \cite{Kang2020}. An optional possibility to be considered is that our X-transition is due to an internal excitation within the $d-d$ states of Ni$^{2+}$ ions \cite{joy1992}. Such, a more trivial origin of our X-transition would classify it along with a similar excitation reported for another layered semiconducting antiferromagnet, MnPS$_3$ \cite{Gnatchenko2011}. The optical transition reported for MnPS$_3$ at $\sim 2.64$~eV is also pretty narrow ($\simeq 0.8$~meV half-width) and appears below the bandgap of this material, though it is identified as an internal transition within the Mn$^{2+}$ ions. Notably, it is active only in the antiferromagnetic phase and displays the magnetic field-induced splitting, suppressed, \textit{notabene} above the spin-flop field, thus largely resembling the properties of the X-transition in NiPS$_3$. The questions, however, arise why the internal transitions within the magnetic ions would be so sharp and, more importantly, why would they appear only in the antiferromagnetic phases (perhaps due to possible change in the crystal field symmetry).

\section{Conclusions}
Concluding we have employed Raman scattering, far-infrared transmission, and photoluminescence spectroscopy measurements, carried as a function of temperature and of the magnetic field, to clarify the controversies regarding the properties of NiPS$_3$ van der Waals antiferromagnet. The observation of the fundamental magnon-gap excitation splitting into two components supports the identification of NiPS$_3$ as a biaxial antiferromagnet which we characterized with the relevant parameters (fundamental magnon-gap energies, $g$-factor, spin-flop field, combination of exchange/anisotropy fields). The overlooked properties of the intriguing optical transition that appear in the near-infrared spectral range have been reported. Our observations, of the magnetic field-induced splitting of this transition in particular, rise pertinent questions about its attribution as a many-body coherent excitonic transition. The reported experimental results call for thorough theoretical works, that can eventually  clarify the origin of the intriguing below-band optical transition in NiPS$_3$. We speculate that similar "excitonic" transitions might be characteristic of other semiconducting antiferromagnets such as, for example, MnPS$_3$.

\section{Acknowledgements}
Numerous valuable discussions with A. Wildes are acknowledged. The work has been supported by the EC Graphene Flagship project. M.P acknowledges support from the Foundation for Polish Science (MAB/2018/9 Grant within the IRA Program financed by EU within SG OP Program).

D.J. performed the data analysis, wrote the preliminary version of the manuscript, and conducted the experiments together with P.K., I.M., D.V., and I.B., whereas M.O., C.F., and M.P. conceptualized the work. All authors discussed the results and contributed to setting the final version of the manuscript.

\providecommand{\noopsort}[1]{}\providecommand{\singleletter}[1]{#1}%

\newpage
\pagenumbering{gobble}

\begin{figure}[htp]
\includegraphics[page=1,trim = 17mm 17mm 17mm 17mm,
width=1.0\textwidth,height=1.0\textheight]{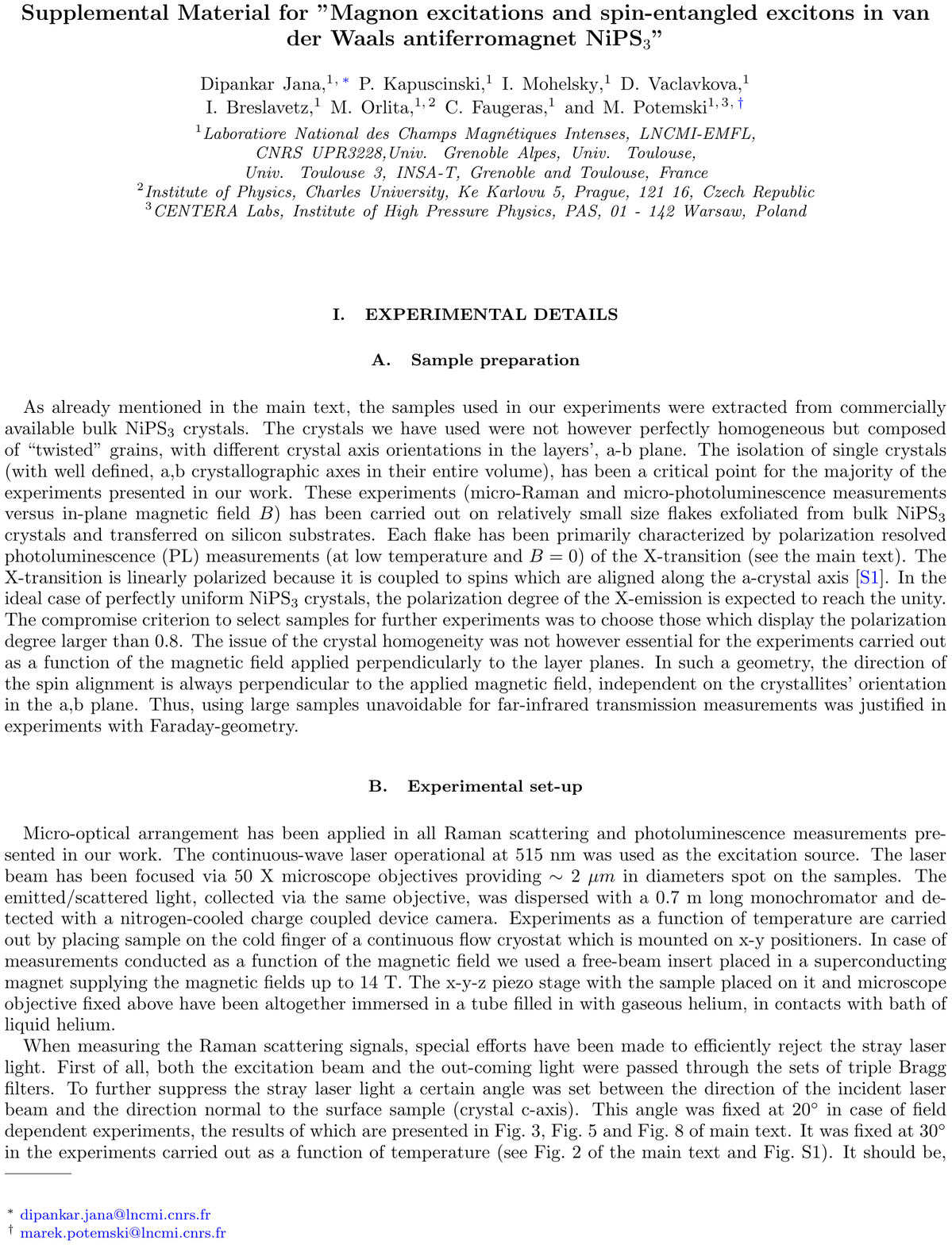}
\end{figure}

\newpage

\begin{figure}[htp]
   \includegraphics[page=2,trim = 17mm 17mm 17mm 17mm,
width=1.0\textwidth,height=1.0\textheight]{NiPS3_Magnon_SM_Dipankar.pdf}
\end{figure}

\newpage

\begin{figure}[htp]
   \includegraphics[page=3,trim = 17mm 17mm 17mm 17mm,
width=1.0\textwidth,height=1.0\textheight]{NiPS3_Magnon_SM_Dipankar.pdf}
\end{figure}

\newpage

\begin{figure}[htp]
   \includegraphics[page=4,trim = 17mm 17mm 17mm 17mm,
width=1.0\textwidth,height=1.0\textheight]{NiPS3_Magnon_SM_Dipankar.pdf}
\end{figure}

\end{document}